\documentstyle[prl,aps]{revtex} \def\narrowtext{} \tighten \twocolumn

\def\be{\begin{equation}}
\def\ee{\end{equation}}
\def\bk{{\bf k}}
\def\bq{{\bf q}}
\def\br{{\bf r}}
\def\da{\downarrow}
\def\ua{\uparrow}
\def\dens{\langle n \rangle}
\def\imchi{{\rm Im}\chi(\bq,\omega)}
\def\comp{(dn/d\mu)}
\def\cs{\chi_s}

\def\rpa{{\small RPA}}
\def\nk{n({\bf k})}

\def\s{\sigma}

\def\ef{\epsilon_F}

\def\tmu{{\tilde\mu}}
\def\nt{N_T(0)}
\def\tc{T_c}
\def\to{\rightarrow}
\def\hb{\hfill\break}
\def\etal{{\it et al.}}

\begin{document}
\draft

\title{Deviations from Fermi-liquid behavior above $T_c$\\
in 2D short coherence length superconductors}
\author{
Nandini Trivedi and Mohit Randeria
}
\address{
Materials Science Division 223,
           Argonne National Laboratory \\
           9700 S.~Cass Ave., Argonne, IL 60439 \\
}
\address{%
\begin{minipage}[t]{6.0in}
\begin{abstract}
We show that there are qualitative differences between the temperature
dependence of the spin and charge correlations
in the normal state of the 2D attractive Hubbard model
using quantum Monte Carlo simulations. The one-particle
density of states shows a pseudogap above $\tc$ with a
depleted $N(0)$ with decreasing $T$.
The susceptibility $\cs$ and the low frequency spin
spectral weight track $N(0)$, which explains the spin-gap scaling:
$1/T_1T \sim \cs(T)$.
However the charge channel is dominated by collective behavior
and the compressibility $dn/d\mu$ is $T$-independent.
This anomalous ``spin-charge separation''
is shown to exist even at intermediate $|U|$ where
the momentum distribution $n(\bk)$ gives evidence for
degenerate Fermi system.
\typeout{polish abstract}
\end{abstract}
\pacs{PACS numbers: 71.27.+a, 74.72.-h, 74.25.-q, 74.40.+k}
\end{minipage}}
\maketitle

\narrowtext

The unusual normal state properties of high $\tc$ superconductors (SC)
have led to many studies exploring possible scenarios for the breakdown of
Fermi liquid theory (FLT) \cite{bedell}.
Much of this effort has been in one of two
directions: search for genuinely new non-Fermi liquid ground states
arising from strong correlations in the proximity of an
insulating state, or search for a low energy scale
such that for temperatures above it one would see deviations from FLT.
In this paper we take a different point of view and ask the question:
Is the normal state of a short coherence length SC
necessarily a Fermi liquid? More specifically, if the ground state
is a condensate of fermion pairs,
and the phase transition leads to a degenerate Fermi system above $\tc$,
do the correlations in that normal state have to be described by Landau's
FLT? If not, what are the characteristic deviations from FLT?

For a weak coupling SC, with the coherence length $\xi_0 \gg a$, the
lattice spacing, the normal state is a FL.
As the coherence length decreases, and $\xi_0/a$ is not a
small parameter, the $\tc$ at which phase coherence is
established and the pair formation scale separate \cite{sre,review}.
Certainly in the opposite extreme of tightly bound pairs,
the state above $\tc$ is a normal Bose liquid.
The question then is whether there is a broad intermediate
coupling regime, especially in 2D, where the normal state has
a ``Fermi surface'' and yet exhibits non-Fermi liquid correlations.

The simplest lattice model within which this problem can be studied
is the attractive Hubbard model where the pair-size can be tuned by
varying the strength of the interaction. In the absence of a small
parameter we use quantum Monte Carlo (MC) simulations \cite{negu.mc,rtms} to
gain insight into the intermediate coupling regime above $\tc$.
While the $U<0$ Hubbard model is not a
realistic microscopic model for the high Tc materials,
it has two merits: its simplicity (fewest parameters) and
the reliability of low temperature MC simulations,
since it is one of the few interacting fermion model which is
free of the ``sign problem'' at all densities \cite{hirsch}.

We begin by summarizing our main results which are obtained in
the normal state ($T > \tc$): \hb
(1) The momentum distribution $\nk$ shows structure reminiscent
of a Fermi surface, though broadened by both thermal and interaction
effects. \hb
(2) A pseudogap opens up in the one-particle density of states (DOS)
well above $\tc$.  The DOS is strongly $T$-dependent -- hence
the notation $\nt$ -- with $d\nt/dT > 0$.
We obtain $\nt$ using a new method which
avoids numerical analytic continuation. \hb
(3) At low $T$, the spin susceptibility $\cs(T) = \nt$, while
the compressibility $\comp$ is $T$-independent. The spin and charge
responses are thus qualitatively different.\hb
(4) The low frequency spectral weight $K(\bq) = \imchi / \omega$
in the spin channel is of the form $K(\bq) \sim \nt/\Gamma_0(\bq)$ where
the $T$-dependence is largely in the DOS and $\Gamma_0(\bq)$ is
essentially the same as in non-interacting system. This naturally
explains the spin-gap scaling $1/T_1T\equiv\sum_q K(\bq)\sim \cs(T)$,
noted in our previous work \cite{rtms}, and
observed in the underdoped
high $T_c$ cuprates \cite{spingap}.

These results show that the normal state of a
short coherence length 2D SC exhibits
marked deviations from usual Fermi liquid behavior.
One obtains a kind of ``spin-charge separation''
with the spin correlations determined by
one-particle excitations, while the charge channel
is dominated by collective excitations.

Consider the attractive Hubbard Hamiltonian
$H = -t\sum_{i,j;\s} c_{i\s}^\dagger c_{j\s} + {\rm h.c.}
- |U| \sum_{i} n_{i\ua}n_{i\da} + \mu \sum_{i;\s}n_{i\s}$,
where the hopping is between nearest neighbor sites.
All energies are measured in units of $t$, and
the lattice spacing $a=1$.

We first establish that in the regime
$|U| = 4$ and $\dens = 0.5$ \cite{cdw} that we focus on here,
our {\it finite system}
results are indeed characteristic of the normal state.
For this we estimate the correlation length $\xi(T)$ from the spatial
decay of the SC order parameter correlation function on
systems of size up to $L=16$.
We find that for $|U| = 4$ we have
$\xi \sim 3-4$  at $T = 1/6$; thus $\xi(T) \ll L$ and the system is normal.
(Details of this analysis and finite size scaling will be presented
elsewhere).

\indent{\bf Single-particle properties:}

To ascertain that the system is degenerate at moderate $|U|$
we study the momentum distribution
$n(\bk) = \sum_\s \langle c_{\bk\s}^\dagger c_{\bk\s} \rangle$
on large lattices of size up to $N=L^2 = 16^2$;
see Fig.~1.  It is clear that $n(\bk)$ shows a rapid variation
with $\bk$ -- a ``Fermi surface'' --
even though it is broadened by the temperature and
also affected by the interactions.
We emphasize that we are {\it not} in the large-$|U|$
preformed boson limit, where the constituent fermions
(tightly bound into singlet pairs) are non-degenerate
and their $n(\bk)$ is independent of $\bk$.

We now turn to the single-particle density of states (DOS)
$N(\omega)$ with $\omega$ measured from $\mu$.
This is given by
$N(\omega) = {1\over N} \sum_{\bk} {\cal A} ({\bk},\omega)$.
Here ${\cal A}$ is the spectral function,
which is related to the
imaginary time Green function $G(\bk,\tau)
= - \langle T [c_{\bk\s}(\tau)c^\dagger_{\bk\s}(0)] \rangle$ via
\begin{equation}\label{gtau}
G({\bk},\tau) = - \int_{-\infty} ^{\infty} d\omega
{{\exp(-\omega\tau)} \over {1+ \exp(-\beta\omega)}} {\cal A} ({\bk},\omega).
\end{equation}
for $0<\tau<\beta=1/T$. To estimate ${\cal A}$ and from
it the DOS, given MC data for $G(\bk,\tau)$,
involves inverting this
Laplace transform.
We avoid the numerical complications inherent in such an analytic
continuation by deriving a general expression
for $N(0)$ in terms of $G(\bk,\tau)$, which is valid {\it provided
there is no low energy scale in the problem}.
Fourier transforming to real space and looking at the local Green
function at $\tau = \beta/2$ we get
$G(\br=0,\beta/2) = -\int d\omega \, {\rm sech}(\beta\omega/2)N(\omega)/2$.
Let $\Omega$ be the frequency scale on which there is structure in the
DOS; for $T \ll \Omega$ we obtain \cite{dos.trick}
\begin{equation}\label{dos}
N(0) \simeq - \beta G(\br=0,\tau=\beta/2)/\pi.
\end{equation}

The one-particle DOS obtained from (\ref{dos}) is plotted in
Fig.~2 as a function of temperature.
We see that for $T>T_c$ a pseudo-gap develops
at the chemical potential and the DOS is depleted
as $T$ is reduced. To emphasize this $T$-dependence
of the DOS we use the notation $\nt$.
This behavior should be compared with that in weak coupling where
the DOS remains featureless above $T_c$, except
for a small fluctuation dip \cite{fluctuations}. The pseudogap
at intermediate coupling may be thought of as the evolution
of the weak coupling fluctuation dip into
a regime where $a/\xi_0$ is no longer a small parameter.

\indent{\bf Spin and charge correlations:} In Fig.~2
we also plot the uniform, static spin susceptibility $\cs$.
The $T$-dependence of $\cs$ was already noted in ref.~\cite{rtms};
what we see here is that this $T$-dependence comes entirely from that
of the one-particle DOS, so that $\cs(T) = \nt$ (to within
the errors inherent in extracting the latter).

It is interesting to ask whether the charge channel also
exhibits the same pseudo-gap. The compressibility $\comp$
was obtained by numerically differentiating \cite{cf.correl}
the average density $\dens$ determined as a function of $\mu$.
We found that $\comp$ shows significant system size dependence
(much more than, e.g., $\cs$); small system data show (finite size)
upturns which are pushed down to lower $T$ with increasing $L$.
The results on the largest lattice ($L = 16$)
plotted in Fig.~3 show that the system
becomes more compressible with increasing $|U|$ (see below).
More significantly, in sharp contrast to the one-particle DOS,
$\comp$ is very weakly $T$-dependent.

Within a simple RPA (p-h bubbles) we obtain
$\comp^\rpa = 2N_0(0)/\left[1 - |U|N_0(0)\right]$.
This is valid only $|U|N_0(0) \ll 1$, where it correctly
explains the trend that attractive interactions increase $\comp$;
for large $|U|$ RPA fails in that it predicts an entirely
spurious instability (phase separation) when $|U|N_0(0) = 1$.
In fact, pairs form at large $|U|$ and their residual interactions
are repulsive, so that the compressibility of the system remains finite
for {\it all} $|U|$. It is simplest to see this in the broken
symmetry state at $T=0$ where we find \cite{compress.0} that,
within mean field (MF) theory, $\comp$ decreases monotonically with $|U|$
from $\comp \simeq 2N_0(0)\left[1 + |U|N_0(0)\right]$ for small-$|U|$ to
$\comp \simeq |U|/4dt^2 - 2/|U|$ for $|U|/t \gg 1$.
The $T=0$ MF result \cite{compress.0}
is also shown in Fig.~3 and is found to be
of the right order-of-magnitude as the normal state MC result;
note that we do not expect $\comp$ to
change dramatically as $T$ goes through $\tc$.

The difference between the spin and charge response functions
could be characterized as a sort of spin charge separation \cite{kr}.
In its mildest form this exists even in
a Landau Fermi liquid where the $\cs$ and $\comp$ are
quantitatively different, the two being renormalized
by different FL parameters: $F_0^a$ and $F_0^s$.
What we see here is much stronger: as a result of strong interactions,
$\cs$ and $\comp$ acquire {\it qualitatively different} $T$-dependences.
As argued above, the spin response is dominated by incoherent
single-particle excitations, which is $T$-dependent because
triplet excitations require breaking up the singlet pair
correlations while the charge channel is dominated
by collective behavior.

\indent{\bf Spin-gap scaling}: We next use our results for
the $T$-dependent DOS $\nt$ to gain insight into the
suppression of low frequency
spectral weight in the spin channel as probed by
$K(\bq) = \lim_{\omega \to 0} \imchi / \omega$.
To contrast with our MC results it may be useful to
recall that for a Fermi liquid (all quantities denoted by
a subscript $0$): $K_0(\bq) = N_0(0)/\Gamma_0(\bq)$
is $T$-independent for $T\ll\ef$
with $\Gamma_0(\bq)\sim v_Fq$.
Further $\sum_\bq\Gamma_0^{-1}(\bq) \simeq N_0(0)$
leads to the Korringa law
$(1/T_1T)_0 = \sum_\bq K_0(\bq) \sim N_0^2(0)$.

In Fig.~4 we plot the MC results for $K(\bq)$ for $\bq \ne 0$
The analytic continuation required to obtain $K$ was done
using the method of ref.~\cite{rtms}.
{}From Fig.~4(a) we see that $K(\bq)$ is
more or less uniformly suppressed at all $\bq$ with
decreasing temperature. Further this $T$-dependence
is similar to that of $\cs(T)$ (or the DOS) shown
at $\bq = 0$ in the Fig.~4(a).
For fixed $T$ the $\bq$-dependence of $K(\bq)$ resembles
that of the non-interacting system as shown in
in Fig.~4(b).

The numerical results thus suggest
that $K(\bq;T) = \alpha\times\nt/\Gamma_0(\bq)$, with
$\alpha$ independent of $T$ and $\bq$, i.e.,
the $T$-dependence comes from a DOS with a pseudo-gap
while the $\bq$-dependence is that of the non-interacting system.
This form for $K(\bq;T)$ leads to
$1/T_1T = \sum_\bq K(\bq;T) \simeq \alpha N_0(0)\nt$, thus
providing a natural explanation for
the spin gap scaling $1/T_1T \sim \cs(T)$ found in our earlier
MC studies \cite{rtms}.

We note that $1/T_1T \sim \cs(T)$ is
observed in the (planar) O  and Y NMR in a large number of
(usually underdoped) cuprates \cite{spingap}.
What distinguishes the results presented here
from spin gap theories \cite{others}
based on spin models, is that the anomalies exist in a system
with itinerant carriers with a large Fermi surface.
Secondly in the present approach these anomalies
are directly related to the $T$-dependence of the one-particle DOS.
On the other hand, since we work with a ``coarse-grained'' model,
rather than a realistic microscopic one,
we cannot discuss the $\bq$-structure of the spin response leading to
differences in the Cu and O NMR, or the doping dependences.
It is also an important open problem to study the normal state
precursors of short coherence length SC with
gap anisotropy and/or nodes.

In conclusion we have shown that the normal state of a 2D
short coherence length SC exhibits characteristic deviations
from Fermi liquid behavior: while the momentum distribution gives
clear evidence for a degenerate Fermi system, the one-particle
DOS shows a pseudo-gap \cite{tmatrix} and the spin and charge
correlations show qualitatively different temperature dependences.

We would like to thank D. Pines, A.J. Leggett, and J.R. Engelbrecht
for very useful suggestions and discussions.
This work grew out of an earlier collaboration with
A. Moreo and R.T. Scalettar whose help is greatly appreciated.
This research was supported by the DOE, Office of Basic
Energy Sciences, Division of Materials Sciences
under contract no.~W-31-109-ENG-38.


\begin{figure}
\caption{
The momentum distribution $n(\bk)$ versus $|\bk|$ along
$[1,0]$ and $[1,1]$ for $|U| = 4$, $\dens = 0.5$ and $T = 0.25$.
The Fermi function ($U = 0$) is shown as the long dashed curve
and the $U= -\infty$ Bose limit result is plotted as the
short dashed curve.
The statistical error bars on all the MC data, unless explicitly shown,
are less than the size of the symbols in this and other figures.
}
\end{figure}

\begin{figure}
\caption{
The one-particle density of states $\nt$ at the chemical potential
(full triangles) and the spin susceptibility $\cs$ (open squares)
as a function of temperature for $|U| = 4$, $\dens = 0.5$ and
$L^2 = 8^2$.
}
\end{figure}

\begin{figure}
\caption{
The compressibility $\comp$ for $\dens = 0.5$, $|U| = 0$ (full line)
and $|U| = 4$ (open circles), as a function of $T$ obtained on a
$16^2$ lattice.  The $T=0$ non-interacting and the $T=0$ mean field result
for $|U| = 4$ are also shown.
}
\end{figure}

\begin{figure}
\caption{
(a) Low frequency spectral weight in the spin channel
$K(\bq;T) = \lim_{\omega \to 0} \imchi / \omega$ for $\bq \ne 0$
(open symbols) plotted along $[1,0]$ and $[1,1]$ for various $T$.
The dashed lines are guides to the eye.
The filled symbols plotted at $\bq = 0$
show $2.0\times \cs(T)$ with $T$ corresponding to that of the open symbols.
Note that the $T$-dependence of $K(q)$ is similar to that of
the susceptibility $\cs(T)$. All of the data is
for $|U| = 4$, $\dens = 0.5$ and $L^2 = 8^2$.
(b) At a fixed $T$ the $\bq$-dependence of $K(q)$ is qualitatively
similar to that of the noninteracting case. To show this we compare the
$K(q)$ at $T=0.25$ (open squares) with the full curve
given by $2\cs(T) K^0(\bq)$, where $K^0(q)$ is
the essentially $T$-independent spectral weight for the
non-interacting system.
}
\end{figure}


\begin{references}

\bibitem{bedell}
See, e.g., {\sl Strongly Correlated Electronic Materials}
edited by K. Bedell {\it el al.}, (Adison Wesley, 1994)
and {\sl High Temperature Superconductivity}
edited by K. Bedell {\it el al.}, (Adison Wesley, 1990).

\bibitem{sre}
C. Sa de Melo, M. Randeria, and J. R. Engelbrecht,
{\sl Phys. Rev. Lett.} {\bf 71}, 3202 (1993).

\bibitem{review}
For a detailed review and references on the BCS-Bose crossover, see
M. Randeria, in {\sl Bose-Einstein Condensation} edited
by A. Griffin, D. Snoke and S. Stringari (Cambridge University Press, 1994).

\bibitem{negu.mc}
(a) R. T. Scalettar \etal, {\sl Phys. Rev. Lett.} {\bf 62}, 1407 (1989);
(b) A. Moreo and D. J. Scalapino, {\sl Phys. Rev. Lett.} {\bf 66}, 946 (1991);
(c) A. Moreo, D. J. Scalapino and  S. R. White, {\sl Phys. Rev. B} {\bf 45},
7544 (1992).

\bibitem{rtms}
M. Randeria, N. Trivedi, A. Moreo, and R. T. Scalettar,
{\sl Phys. Rev. Lett.} {\bf 69}, 2001, (1992).

\bibitem{hirsch}
J. E. Hirsch, {\sl Phys. Rev. B} {\bf 28}, 4059 (1983).

\bibitem{spingap}
W. W. Warren \etal, {\sl Phys. Rev. Lett.} {\bf 62}, 1193 (1989);
M. Takigawa \etal, {\sl Phys. Rev. B} {\bf 43}, 247 (1991);
H. Alloul {\it et al.}, {\sl Phys. Rev. Lett.} {\bf 70}, 1171 (1993);
R. E. Walstedt {\it et al.}, {\sl Phys. Rev. Lett.} {\bf 72}, 3610 (1994).

\bibitem{cdw} We choose $\dens \ne 1$ since
we are not interested in the
competition with the CDW state at half-filling.
See: R. Micnas, J. Ranninger and S. Robaskiewicz,
{\sl Rev. Mod. Phys.} {\bf 62}, 113 (1990).

\bibitem{dos.trick}
For low frequencies
$N(\omega) = N(0) + N^\prime(0)\omega + \ldots$,
where $N^{(m)}(0) \sim N(0)/\Omega^m$.
Substituting this in the preceding equation we obtain
(\ref{dos}) with corrections of order $(T/\Omega)^2$.

\bibitem{fluctuations}
C. Di Castro, C. Castellani, R. Raimondi and A. Varlamov,
{\sl Phys Rev.} {\bf B  42}, 10211 (1990). For the resulting
effects on spin correlations, see: M. Randeria and A. Varlamov,
{\sl Phys Rev.} {\bf B 42}, 10211 (1994) and D. Rainer
(private communication).

\bibitem{cf.correl}
We could not use
$\langle n_i n_j \rangle - \langle n_i \rangle\langle n_j \rangle$
which is prone to errors due to large cancellations.

\bibitem{compress.0}
{}From an extension of the $T=0$ analysis by
L. Belkhir and M. Randeria, {\sl Phys. Rev.} {\bf B 45},
5087 (1992) and {\sl Phys. Rev.} {\bf B 49}, 6829 (1994),
we find $\comp = (\partial n/\partial \tmu)/
\left[1 - |U|(\partial n/\partial \tmu)/2\right]$
with $\partial n/\partial \tmu = \Delta^2\sum_\bk E^{-3}
+ \left(\sum_\bk \xi E^{-3}\right)^2/ \left(\sum_\bk E^{-3}\right)$.
Here $\tmu = \mu + \dens |U|/2$, and the rest of the
notation is standard: $\xi_\bk$ is the energy measured from $\tmu$,
$\Delta$ is the gap, and $E_\bk$ is the Bogoliubov quasiparticle energy.

\bibitem{kr}
The parallels between spin charge separation (SCS)
in the resonating valence bond picture and in the BCS ground state have
been discussed by
S. A. Kivelson and D. S. Rokhsar,
{\sl Phys. Rev.} {\bf B 41}, 11693 (1990).
Here we find that pairing correlations alone can also give rise to SCS
without any broken symmetry.

\bibitem{others}
A. J. Millis and H. Monien,
{\sl Phys. Rev. Lett.} {\bf 70}, 2810 (1993),
{\bf 71}, 210 (1993) (E);
A. Sokol and D. Pines,
{\sl Phys. Rev. Lett.} {\bf 71}, 2813 (1993).

\bibitem{tmatrix}
After this work of was completed we learned from R. Micnas
of self consistent T-matrix calculations at low densities by
J. J. Rodriguez-Nunez \etal (unpublished)
which also find a pseudo-gap in the DOS.

\end{references}
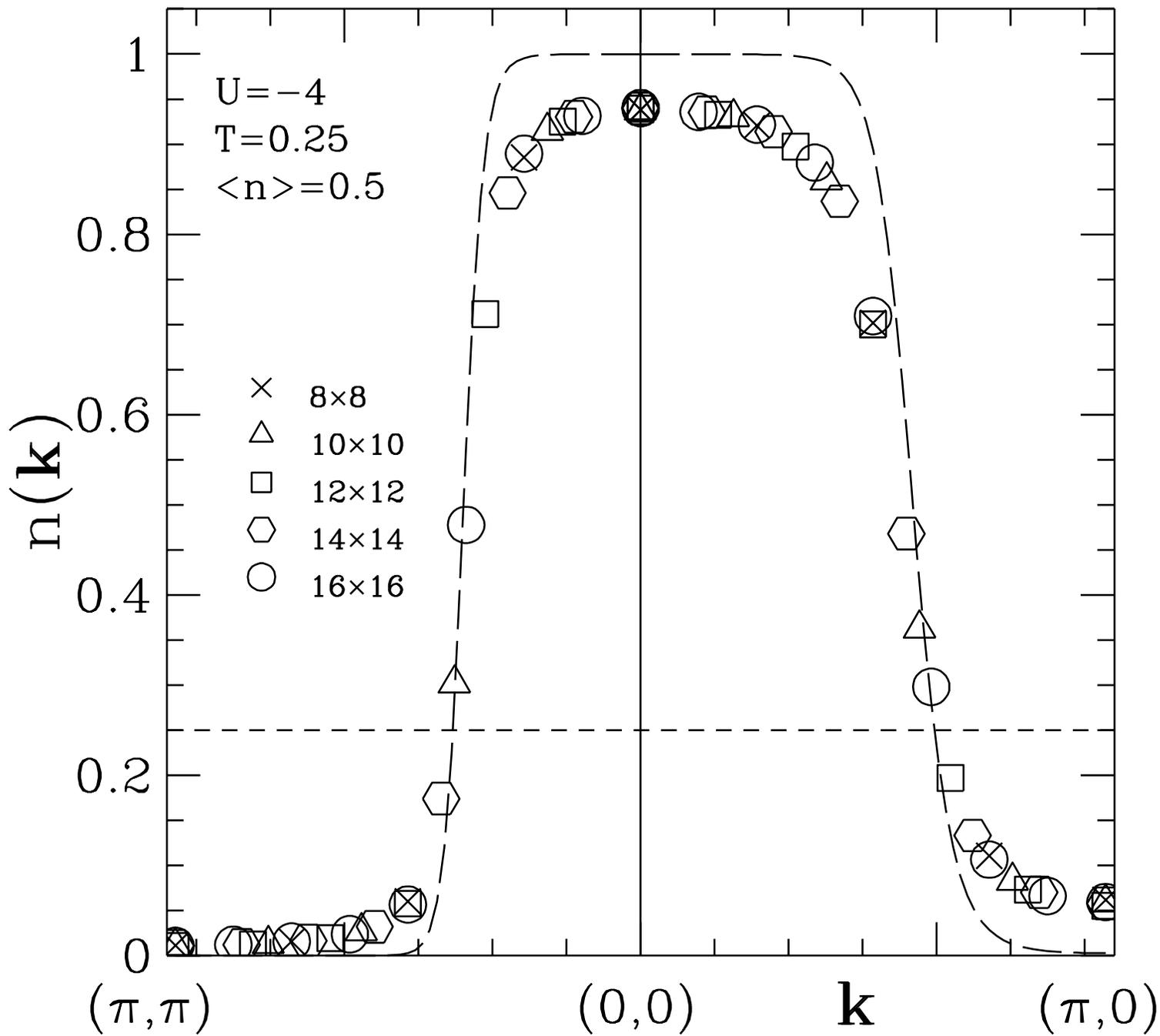
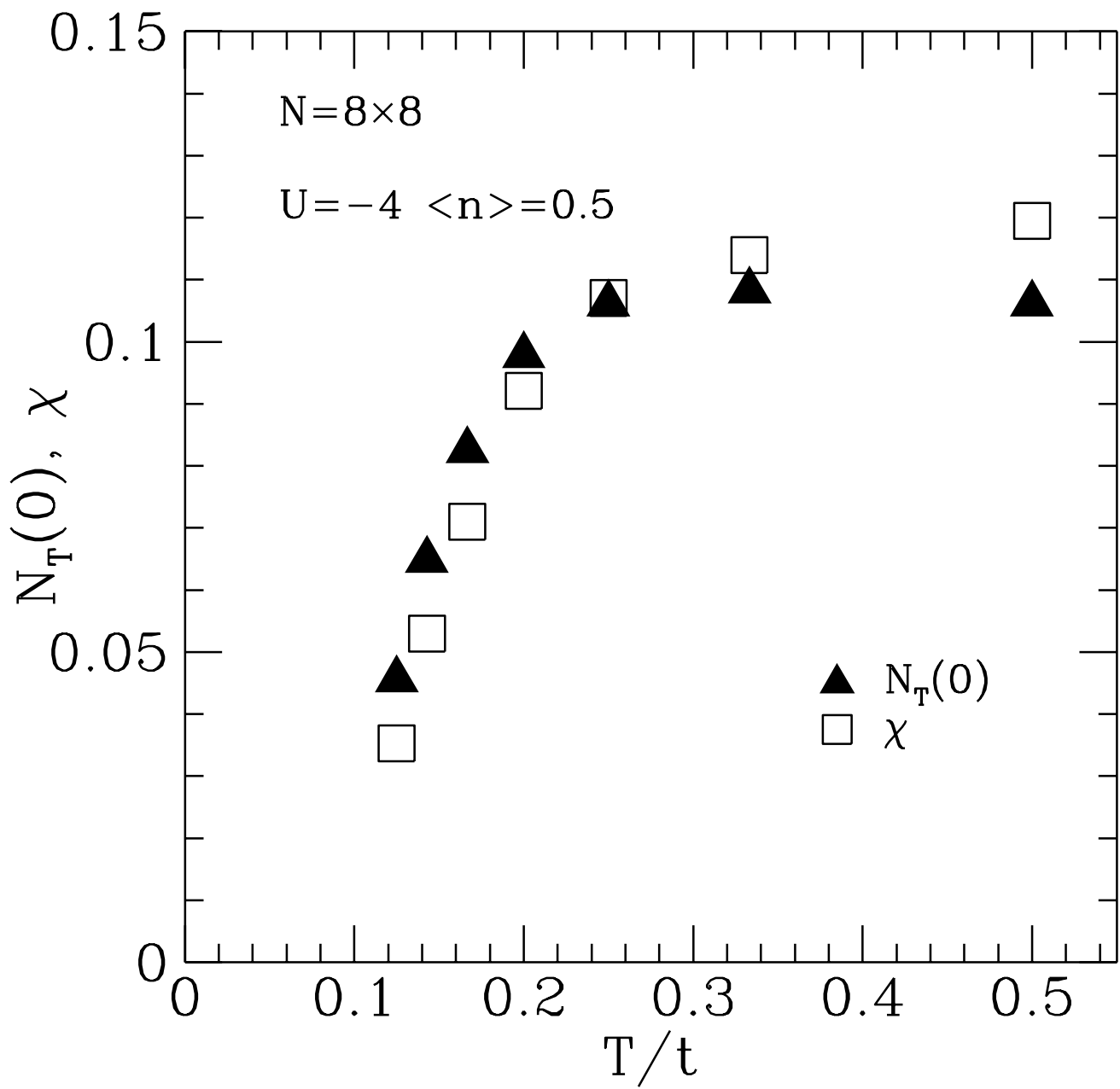
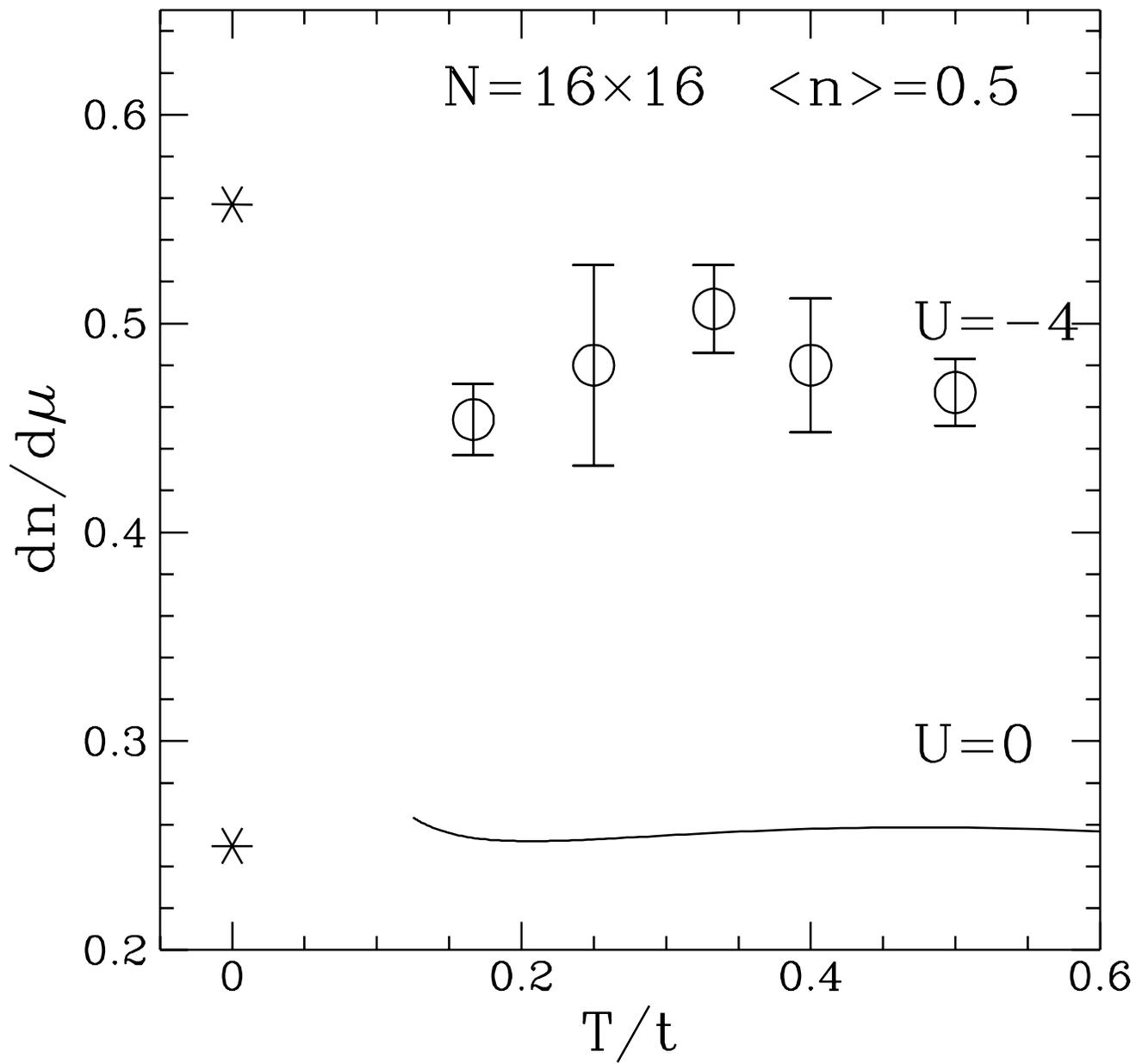
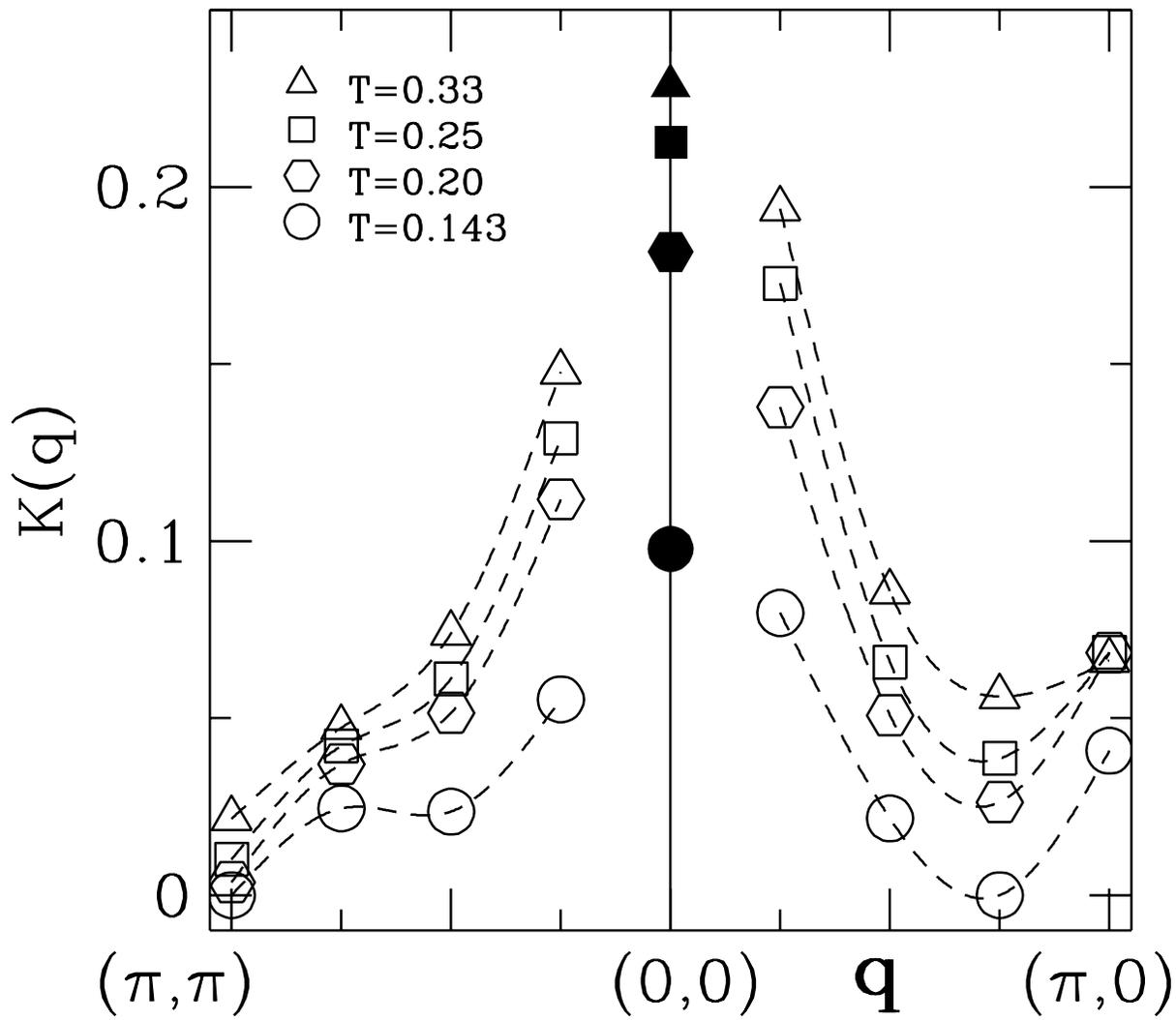
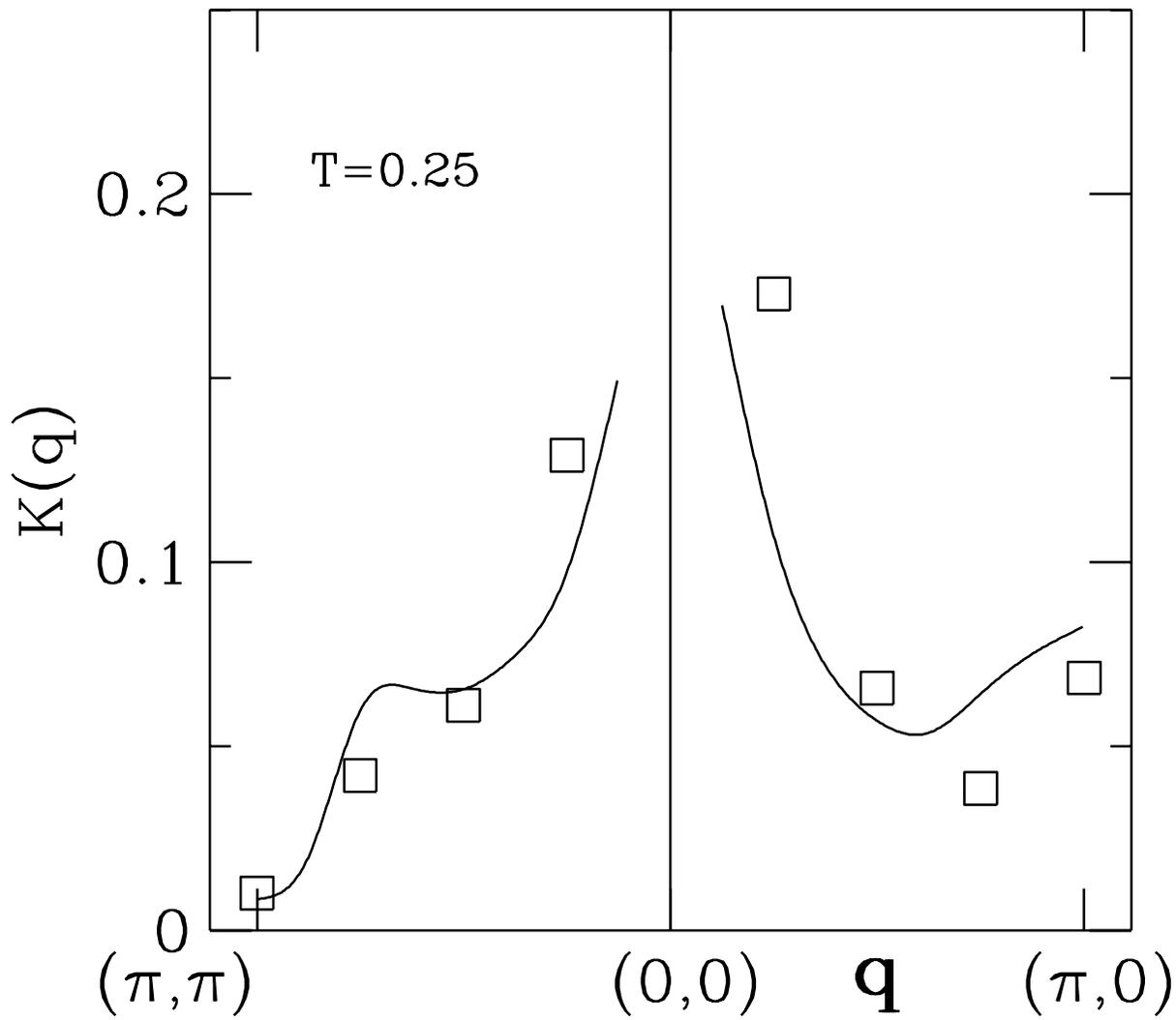
\end{document}